\documentclass[aps,pra,twocolumn,superscriptaddress]{revtex4-2}
\usepackage[utf8]{inputenc}
\usepackage{soul}
\usepackage{textcomp}
\usepackage[english]{babel}
\usepackage{graphicx}
\usepackage{dcolumn}
\usepackage{bm}
\usepackage{amsmath}
\usepackage{verbatim}     
\usepackage[dvipsnames]{xcolor}       
\usepackage[normalem]{ulem}   
\usepackage{bbold}
\usepackage{mathrsfs}
\usepackage[mathscr]{eucal}
\usepackage{physics}
\usepackage{algorithmic}

\newcommand{\qo}[1]{``#1''}                               		
\newcommand{\beq}{\begin{equation}}
\newcommand{\eeq}{\end{equation}}
\newcommand{\bei}{\begin{itemize}}			
\newcommand{\eei}{\end{itemize}}			

\usepackage{hyperref}
\hypersetup{
   pdfpagemode=None,
   pdfstartpage=1,
   pdfmenubar=true,
   pdftoolbar=true,
   colorlinks = true,
   linkcolor=blue,
   citecolor=blue,
   urlcolor=blue,
   bookmarksopen=false
 }

\begin{document}
\title{Retrieving space-dependent polarization transformations via near-optimal quantum process tomography}


\author{Francesco Di Colandrea}
\email{francesco.dicolandrea@unina.it}
\affiliation{Dipartimento di Fisica, Universit\`{a} di Napoli Federico II, Complesso Universitario di Monte Sant'Angelo, Via Cintia, 80126 Napoli, Italy}
\author{Lorenzo Amato}
\thanks{FDC and LA contributed equally to this work.}
\affiliation{Condensed Matter Theory Group, Paul Scherrer Institute, CH-5232 Villigen PSI, Switzerland}
\affiliation{Laboratory for Solid State Physics, ETH Zurich, CH-8093 Zurich, Switzerland}
\author{Roberto Schiattarella}
\affiliation{Dipartimento di Fisica, Universit\`{a} di Napoli Federico II, Complesso Universitario di Monte Sant'Angelo, Via Cintia, 80126 Napoli, Italy}
\author{Alexandre Dauphin}
\affiliation{ICFO -- Institut de Ciencies Fotoniques, The Barcelona Institute of Science and Technology, 08860 Castelldefels (Barcelona), Spain}
\author{Filippo Cardano}\email{filippo.cardano2@unina.it}
\affiliation{Dipartimento di Fisica, Universit\`{a} di Napoli Federico II, Complesso Universitario di Monte Sant'Angelo, Via Cintia, 80126 Napoli, Italy}
\begin{abstract}
An optical waveplate rotating light polarization can be modeled as a single-qubit unitary operator, whose action can be experimentally determined via quantum process tomography. Standard approaches to tomographic problems rely on the maximum-likelihood estimation, providing the most likely transformation to yield the same outcomes as a set of experimental projective measurements. The performances of this method strongly depend on the number of input measurements and the numerical minimization routine that is adopted. Here we investigate the application of genetic and machine learning approaches to this problem, finding that both allow for accurate reconstructions and fast operations when processing a set of projective measurements very close to the minimal one. We apply these techniques to the case of space-dependent polarization transformations, providing an experimental characterization of the optical action of  spin-orbit metasurfaces having patterned birefringence. Our efforts thus expand the toolbox of methodologies for optical process tomography. In particular, we find that the neural network-based scheme provides a significant speed-up, that may be critical in applications requiring a characterization in real-time. We expect these results to lay the groundwork for the optimization of tomographic approaches in more general quantum processes, including non-unitary gates and operations in higher-dimensional Hilbert spaces. 
\end{abstract}

\maketitle
\section{Introduction}
Quantum process tomography (QPT) consists in determining all the parameters characterizing a quantum evolution from a set of experimental measurements~\cite{Chuang1997}. This represents a crucial task for most quantum applications. If the process under investigation is a genuine \emph{quantum black box}, i.e., no preliminary information is available, QPT unveils the mathematical structure of the unknown transformation, thus enabling the prediction of how a specific input state will change under its action. On the other hand, QPT is key to certifying if a quantum device is properly working, for instance, validating the security of a quantum communication channel~\cite{Bouchard2019} or reconstructing a transfer matrix within a quantum circuit~\cite{Goel2022}. In the past decades, this technique has been employed to characterize different quantum architectures, such as nuclear magnetic resonances~\cite{PhysRevA.64.012314}, atoms in optical lattices~\cite{PhysRevA.72.013615}, trapped ions~\cite{PhysRevLett.92.220402,PhysRevLett.97.220407}, superconducting circuits~\cite{PhysRevB.82.184515,Bialczak2010} and photonic setups~\cite{PhysRevLett.91.120402,Altepeter2003,PhysRevLett.93.080502,Lobino2008, Bongioanni2010,Rahimi-Keshari2013,Ndagano2017,Anton2017,PhysRevA.98.052327}. 

In the simplest picture of linear and unitary processes, in principle QPT could be accomplished by extracting analytical relations between the transformation parameters and the outcomes of suitable projective measurements~\cite{LeRoy-Brehonnet1997,Laing2012}. However, realistic experimental noise compromises the feasibility of such approach, typically yielding non-physical results. This limitation can be overcome by mapping the reconstruction of the process into an optimization problem, as first proposed for the tomography of quantum states~\cite{Hradil1997,Tan1997,Banaszek1999,Hradil2000,James2001,Rehacek2001,Kaznady2009}. 
The gold standard is the \emph{maximum-likelihood} approach \cite{PhysRevLett.93.080502}, consisting of the minimization of the negative log-likelihood. The latter represents a notion of distance between a set of experimental outcomes 
and the corresponding theoretical predictions, based on the estimated quantum process.

In this framework, the most elementary scenario is the characterization of an SU(2) gate $U$ acting on a two-level system, encoding a qubit of quantum information. In photonic setups, qubits can be encoded into optical polarization, with $U$ implemented via one or multiple birefringent waveplates. Accordingly, the characterization of devices acting on  light polarization can be nicely solved within the mathematical paradigm of QPT~\cite{Aiello2006,Jones41}. In this analysis, we will not consider the broader subject of depolarizing optical transformations that are described using the Mueller matrix formalism. This will be addressed in a subsequent study.

While standard waveplates rotate uniformly the polarization of a light beam, several photonic applications involve light propagation through inhomogeneous devices or materials that implement space-dependent polarization transformations, possibly exhibiting a complex spatial structure. These applications include polarization imaging~\cite{Solomon1981} and multiplexing~\cite{Davis2005}, or the generation and manipulation of structured light~\cite{Rubinsztein-Dunlop2017,Forbes2021,Piccardo2022} through spin-orbit metasurfaces~\cite{Neshev2018}. In these cases, reconstructing the whole transformation requires the determination of a two-dimensional (2D) spatially-varying unitary operator $U(x,y)$, which in turn implies that the QPT procedure is iterated over multiple positions. Performing the full tomography by relying upon \emph{brute-force} minimization routines may become significantly lengthy when increasing the number of iterations, depending on the desired spatial resolution and the system size. 

Here we investigate the optimization of numerical techniques assisting the tomography of SU(2) gates, experimentally encoded into unitary polarization rotations. By optimization, we mean achieving a satisfactory level of accuracy in the characterization process while using minimal resources, including the number of measurements and the time consumed by the numerical routines.

Considering a collimated light beam propagating along the $z$ axis, we aim at determining the parameters of a space-dependent polarization rotation $U(x,y)$ implemented by a thin birefringent optical element, lying in a transverse plane $z=z_0$. We specifically consider two complementary approaches to the reconstruction of $U$ at each transverse position $(x,y)$, namely a genetic algorithm (GA) and supervised machine learning (ML). In the first case, we devise a genetic search of the minimum of the cost function, emulating the principle of Darwinian natural selection~\cite{holland,Whitley1994,mitchell1998,Schmitt2001}. GAs have been largely employed in the contexts of quantum simulation~\cite{GAquantum1}, quantum annealing~\cite{Hegde2022}, as well as for diverse optical applications~\cite{Mahlab91,Kihm96,Woodward2016,Woodward17,Michaeli2018,Spagnolo2017,Pu2020,Bielak2021,Karpov2022}. The second approach relies on training a neural network~(NN) to predict the process generating a limited number of experimental outcomes. Nowadays, ML is ubiquitous in physics~\cite{Jordan2015}, playing a special role in quantum applications~\cite{review_alex}. Photonic setups represent an important testbed to probe its potentialities~\cite{Genty2021}. 

We first test the algorithms with synthetic experimental data, generated via numerical simulations carried out for single unitary processes.  In the context of space-dependent transformations, this would correspond to the polarization rotation associated with a sufficiently small spatial region.  By keeping the number of input measurements to six, we compare the performances of these approaches with the popular \emph{NMinimize} routine from Mathematica~\cite{nminimize}, taking into account both the timing and the accuracy of the prediction.  Finally, we adapt these routines to characterize the entire polarization transformation generated by chosen combinations of inhomogeneous optical waveplates, such as liquid-crystal (LC) metasurfaces \cite{Rubano2019,DErrico2020}.

\section{Polarization rotations as SU(2) processes}\label{sec:su2}
\begin{figure*}
    \centering
    \includegraphics[width=0.83\linewidth]{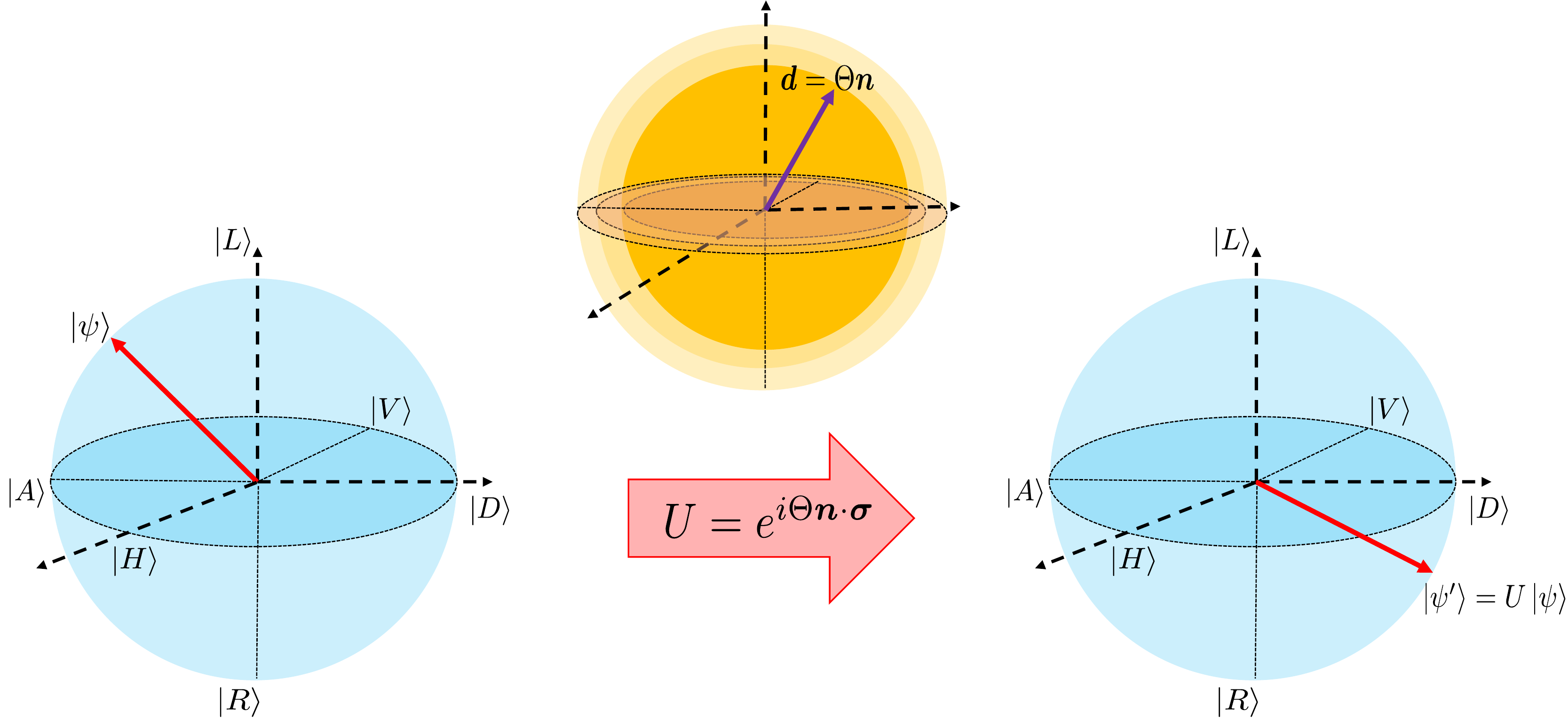}
    \caption{Geometric representation of the parametrization $\left(\Theta, \bm{n} \right)$ of SU(2) gates. Polarization qubits are represented as points on the Poincaré sphere (cyan sphere). Two qubits are connected via an SU(2) map $U$, $\ket{\psi'}=U\ket{\psi}$, represented as a point at $\bm{d}=\Theta \bm{n}$, lying on a sphere of radius $\Theta$ (yellow sphere).}
    \label{fig:SU2}
\end{figure*}

Within Jones' formalism, the polarization state of a fully-polarized light beam, or a single photon, can be expressed as a two-component complex vector $\psi=(\alpha,\beta)^T$, where $|\alpha|^2+|\beta|^2=1$. Here and in the following, we are considering left and right circular polarizations as basis states $(1,0)^T$ and $(0,1)^T$, respectively. Building on a one-to-one correspondence between this representation and the ket notation for quantum states, we label circular basis states as $\ket{L}$ and $\ket{R}$, respectively, and represent $\psi$ as a qubit $\ket{\psi}=\alpha \ket{L}+\beta\ket{R}$, that can be visualized as a point on the unit-radius Bloch (or Poincaré) sphere [see Fig.\ \ref{fig:SU2}]. Within this formalism, the action of a unitary rotation of light polarization, or equivalently a single-qubit unitary gate, is captured by an SU(2) operator. This can be expressed as
\begin{equation}
    U=\exp{-i\Theta(\bm{n}\cdot\bm{\sigma})},
    \label{eq:exprep}
\end{equation}
whose matrix form reads 
\begin{equation}
U=\begin{pmatrix}
\cos \Theta -i \sin \Theta \,n_z && -i\sin \Theta \,(n_x-i n_y)\\
-i\sin \Theta \,(n_x+i n_y) && \cos \Theta + i \sin \Theta \,n_z
\end{pmatrix}.
\label{eqn:su2matrix}
\end{equation}
In Eqs.\ \eqref{eq:exprep}-\eqref{eqn:su2matrix}, $\Theta\in[0,\pi]$, $\bm{n}=(n_x,n_y,n_z)$ is a real-valued unit-norm vector and $\bm{\sigma}=\left({\sigma_x,\sigma_y,\sigma_z}\right)$ is a vector whose components are the three Pauli operators. 

This representation allows one to map any SU(2) matrix into a point on a sphere of radius $\Theta$, whose direction is given by $\bm{n}$. 
In this picture, the action of the operator defined in Eq.\ \eqref{eqn:su2matrix} corresponds to a counterclockwise rotation of the Poincaré sphere around the ${\bm{n}\text{-axis}}$ through an angle $2\Theta$ [see Fig.~\ref{fig:SU2}]. Besides the operator in Eq.~\eqref{eqn:su2matrix}, a generic polarization rotation would also include a global phase $\Phi$, which we are not considering here. Its possible role is discussed in detail below. 

From the SU(2) group property, it follows that the cascaded action of multiple waveplates can also be written as in Eq.~\eqref{eqn:su2matrix}. In this representation, the unit vector $\bm{n}$ gives the position on the Poincaré sphere of the polarization eigenstates associated with the optical transformation. These are optical modes whose polarization state is not altered after passing through the optical sequence, as these simply acquire a phase factor $e^{\pm i\Theta}$. 

In the tomographic reconstruction of $U$, we aim at determining the four parameters $\left(\Theta,\bm{n}\right)$ describing the whole transformation, by investigating how certain input polarizations are modified when passing through the chosen optical elements. This information is retrieved via projective measurements, illustrated in the next section. 

\section{Projective measurements}
\label{sec:polarimetricmeas}
The basic idea behind the tomographic approach is that each projective measurement conceals some information about the transformation parameters. To exploit this idea, we repeatedly prepare a sequence of $N_\text{in}$ input polarization states $\ket{\psi_\text{in}^i}$ ($i=1,...,N_\text{in}$), let them evolve under the action of $U$, and project them on a sequence of $N_\text{out}$ states $\ket{\psi_\text{out}^j}$ ($j=1,...,N_\text{out}$). Finally, we use a power meter to record the optical power
\begin{equation}
    I_{ij}=I_0\abs{\mel{\psi_\text{out}^j}{U}{\psi_\text{in}^i}}^2,
    \label{eq:polar_mes}
\end{equation}
where $I_0$ is the total power of the beam, and we are assuming that both $U$ and the detection process are not affected by losses.

As anticipated above, it is clear that global phase factors $\Phi$ in the unitary map $U$ do not affect any measurement of the form of Eq.~\eqref{eq:polar_mes}. Accordingly, any tomography based only on projective measurements cannot detect them. While this is not an issue in the case of spatially-uniform rotations, it could limit our ability to characterize entirely the action of space-dependent transformations where also $\Phi$ is inhomogeneous. If needed, this quantity could be obtained with standard interferometric measurements. Our ignorance of global phase factors also implies that an SU(2) process $U$ with parameters $(\Theta,\bm{n})$ cannot be distinguished from the one corresponding to $(\pi-\Theta,-\bm{n})$, since the latter is simply $-U$ \cite{Flaschner2016,Li2016,Tarnowski2019,Yi2023}. The implications of this ambiguity will be  
further handled when analyzing complex polarization transformations in Sec.~\ref{complex}. 

Our projective measurements involve the six polarization states forming the mutually unbiased bases of SU(2): $\ket{L}$ and $\ket{R}$, $\ket{H}=\left(\ket{L}+\ket{R}\right)/\sqrt{2}$ and $\ket{V}=\left(\ket{L}-\ket{R}\right)/\sqrt{2}$ (horizontal and vertical polarizations, respectively), and $\ket{D}=\left(\ket{L}+i\ket{R}\right)/\sqrt{2}$ and $\ket{A}=\left(\ket{L}-i\ket{R}\right)/\sqrt{2}$ (diagonal and antidiagonal polarizations, respectively). 
Considering other sets of states, as shown for instance in Ref.\ \cite{QST4proj} in the case of quantum state tomography of polarization qubits, is out of the scope of the present work. We plan to consider this aspect in the future for further optimization of the proposed techniques.

\section{OPTIMIZED PROCESS TOMOGRAPHY}

The set of equations \eqref{eq:polar_mes} contains analytical relations between the parameters $(\Theta,\bm{n})$ and the projective measurements $I_{ij}$. As such, these could be solved exactly or numerically in order to reconstruct the unknown process. Although licit, this straightforward approach proves to be highly unreliable because of the uncorrelated experimental noise between different polarimetric measurements, which often entails the prediction of non-physical results (for example, the parameter $\Theta$ or some components of the vector $\bm{n}$ feature an imaginary part) \cite{Aiello2006}. This inconvenience can be avoided by casting the tomography as an optimization problem, which allows retrieving the ``most likely" SU(2) process compatible with the experimental results. The reconstruction is thus accomplished by minimizing the distance between a set of experimental outcomes and the corresponding theoretical predictions.

\subsection{The minimization approach}
The ``distance" mentioned above is quantitatively estimated in terms of a cost function, whose minimization is typically achieved via automatized  routines. 

Assuming that the light detection system is only affected by Gaussian noise, the log-likelihood cost function is expressed as \cite{James2001}
\begin{equation}
\mathcal{L}=\sum_{\alpha} \frac{(I^\text{th}_{\alpha}-I^\text{exp}_{\alpha})^2}{I^\text{th}_{\alpha}},
\label{eqn:likelihood}
\end{equation}
where $\alpha$ runs over the chosen input-output combinations of Eq.~\eqref{eq:polar_mes}. For each projective measurement, $I^\text{th}_{\alpha}$ and $I^\text{exp}_{\alpha}$ represent the  theoretical and experimental normalized light intensities, respectively. The normalization is performed with respect to the total optical power $I_0$ \footnote{Equation~\eqref{eqn:likelihood} is essentially the extension to coherent light beams of the formula originally provided in Ref.~\cite{James2001}.}.  

To keep both experimental and computational time at a minimum, one would consider a reduced set of projective measurements to be carried out and processed. However, decreasing the amount of data comes at the expense of the accuracy of the tomography. In Appendix~\ref{app:supp_proof}, we prove that 
a minimum of five measurements is needed to reconstruct a generic SU(2) operator. Here we focus on the near-optimal case corresponding to six of these, as it proved more robust to experimental noise. Even in this case, we find that process tomography based on GAs and NNs reconstructs SU(2) matrices with significantly large fidelities (on average). Benchmarking these results against those obtained through all available minimization routines is out of the scope of the present work. Among common choices, we employ \emph{NMinimize} (NMin) from Mathematica~\cite{PhysRevLett.93.080502,James2001,Bongioanni2010}. This routine embeds several optimization methods. By default, it automatically picks the most convenient one based on the type of cost function~\cite{nminimize}.

\subsection{The genetic algorithm approach}
Our genetic algorihtm evolves real-valued \emph{individuals} (or \emph{chromosomes})
\begin{equation}
    \bm{x} = (\Theta,n_x,n_y,n_z).
    \label{eq:individual}
\end{equation}
Each individual is a candidate to provide an optimal parametrization of the unknown process in terms of $\Theta$ and $\bm{n}$. By means of operators mimicking the natural selection mechanism, the GA selects for reproduction those individuals better minimizing the Mean Squared Error (MSE) cost function 
\begin{equation}
\mathcal{L}_{\text{MSE}}=\sum_{\alpha} (I^\text{th}_{\alpha}-I^\text{exp}_{\alpha})^2
\label{eqn:MSE}
\end{equation}
within the current generation. Their genetic inheritance is transmitted to the next generations, thus driving future populations toward the optimal solution. In our implementation, the reproduction occurs in the form of \emph{blend crossover}~\cite{eshelman1993real}.
Furthermore, to explore a wider region of the parameter landscape, genetic mutations are included in our workflow in the form of Gaussian noise on single genes. 
The maximum number of generations is used as a termination criterion. The complete sequence of operators implementing our genetic reconstruction is detailed in Appendix \ref{app:GAtomography}.

\subsection{The neural network approach}
Neural networks constitute our second alternative. 
We apply supervised learning to train a feedforward NN to predict the optimal $(\Theta,\bm{n})$ parametrization reproducing a set of experimental outcomes. To perform QPT with our NN, the training set consists of a suitable set of projective measurements, as prescribed by Eq.~\eqref{eq:polar_mes}, while the outputs are related to the transformation parameters $\left(\Theta,n_x,n_y\right)$, with $n_z$ obtained from the normalization condition. Interestingly, the learning procedure is proficiently achieved only when using  half of the sphere associated with the SU(2) space [see Fig.\ \ref{fig:SU2}], restricting the training samples to the northern hemisphere $n_z>0$ (or equivalently the other hemisphere). This is necessary to remove the ambiguity between $U$ and $-U$ discussed in Sec.~\ref{sec:polarimetricmeas} \cite{Flaschner2016,Li2016,Tarnowski2019,Yi2023}. 

Starting from a random initialization, the learning process dynamically refines all the weights of the NN via the optimization of a cost function (\emph{loss function}), measuring the distance between the prediction with the current settings and the correct outputs (\emph{labels}).
The learning process is divided into different time blocks (\emph{epochs}), and it is completed when the cost function converges to a global minimum. The MSE function
is chosen as a cost function to be minimized during supervised learning.
Further details on the structure of the NN and the learning phase are reported in Appendix~\ref{app:MLtomography}. 
\section{NUMERICAL EXPERIMENTS} 
To validate our methods for reconstructing a generic SU(2) transformation, we carry out a series of numerical experiments, comparing the performance of the GA and ML approaches with the minimization of the likelihood function in Eq.\ \eqref{eqn:likelihood} performed with NMin, when processing $N=6$ measurements. We refer to Appendix~\ref{app:hyper-parameters} for the explicit declaration of all hyper-parameters used in our routines. 

First, we build a set of $10^3$ random unitary operators, sampling uniformly the SU(2) space. Then, as discussed in Sec.~\ref{sec:polarimetricmeas}, we compute the outcomes of a set of selected projective measurements (further details below). We also study the effect of various levels of disorder on the synthetic experiments. This is introduced as a zero-mean Gaussian noise with increasing standard deviation $\Delta$ affecting the angular settings of the optical waveplates used to realize each projective measurement, thus mimicking non-ideal optical measurements.  
The chosen input-output combinations are  
\begin{equation}
\begin{split}
I_{LL}&=n_z^2 \sin ^2(\Theta)+\cos ^2(\Theta),\\
I_{HH}&=n_x^2 \sin ^2(\Theta)+\cos ^2(\Theta),\\
I_{LH}&=\frac{1}{2}\left(1+2 n_x n_z \sin^2 (\Theta)+n_y \sin (2 \Theta)  \right),\\
I_{LD}&=\frac{1}{2}\left(1+2 n_y n_z \sin^2 (\Theta )-n_x \sin (2 \Theta)  \right),\\
I_{HL}&=\frac{1}{2}\left(1+2 n_x n_z \sin^2 (\Theta )-n_y \sin (2 \Theta)  \right),\\
I_{HD}&=\frac{1}{2}\left(1+2 n_x n_y \sin^2 (\Theta)+n_z \sin (2 \Theta)  \right),
\end{split}
\label{eqn:expdata}
\end{equation}
where $I_{ij}= \abs{\mel{j}{U}{i}}^2/I_0$ denotes the normalized light power recorded when we prepare the polarization state $\ket{i}$ and project onto $\ket{j}$ after the evolution $U$.

\begin{figure*}
    \includegraphics[width=0.98\linewidth]{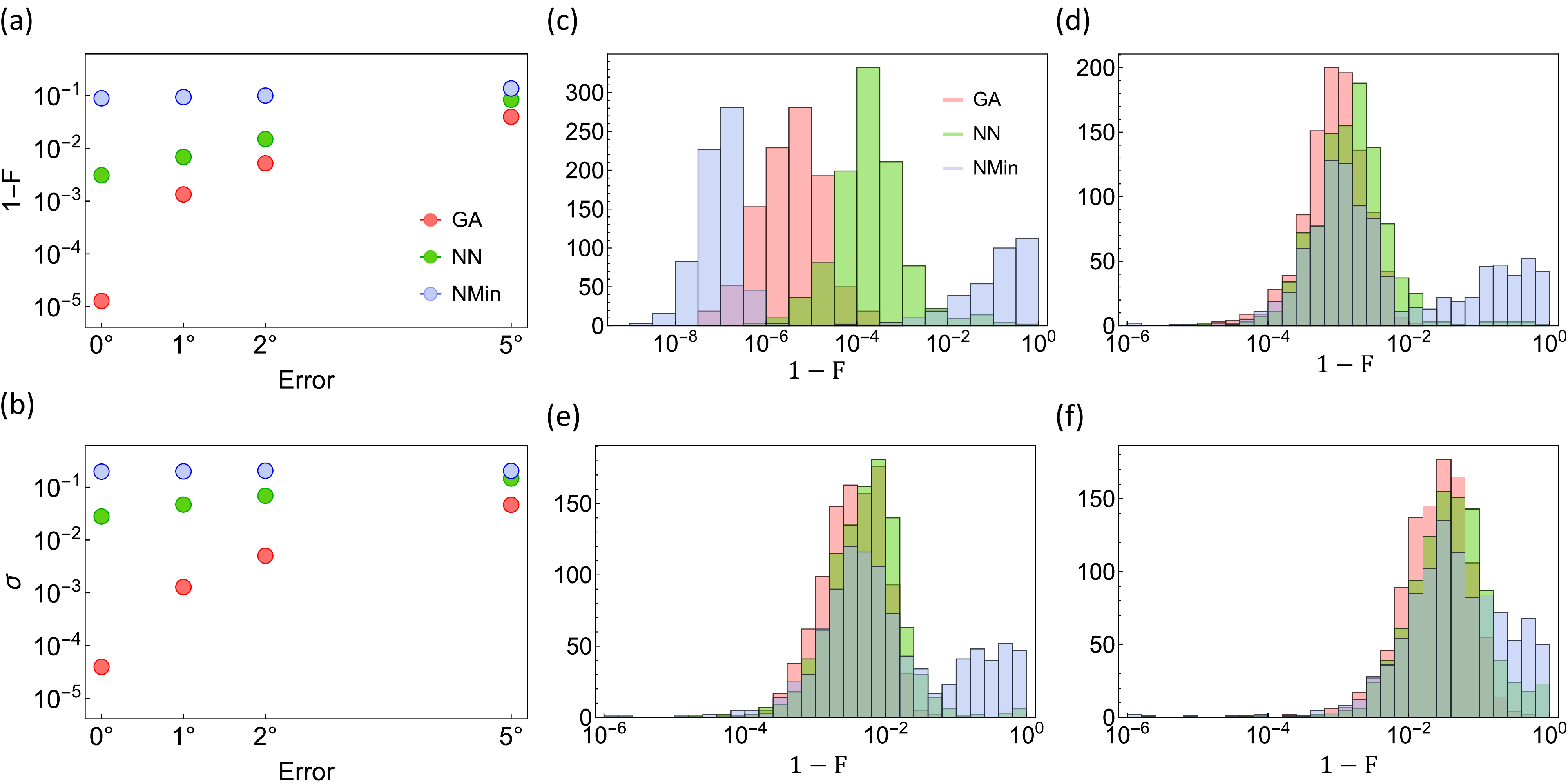}
    \caption{(a) Log-plot of the average infidelities of the three optimization algorithms (see legend) for different levels of Gaussian noise. The average is computed over $10^3$ numerical experiments. (b) Log-plot of the standard deviations of the fidelity distributions for the same synthetic realizations as in panel (a). (c-f) Infidelity distributions are represented as histograms for different levels of noise: (c) $\Delta=0^\circ$, (d) $\Delta=1^\circ$, (e) $\Delta=2^\circ$, (f) $\Delta=5^\circ$.
    }
    \label{fig:infid}
\end{figure*}
In order to quantify the agreement between the theoretical model and the prediction resulting from each reconstruction algorithm, we compute the fidelity~\cite{PhysRevA.71.062310,Wang2009,Cabrera2011}
\begin{equation}
F=\frac{1}{2}\,\biggl|\Tr(U_\text{th}^{\dagger}U_\text{exp})\biggr|,
\end{equation}
where $U_\text{th}$ and $U_\text{exp}$ are the target and predicted processes, respectively. We note that the fidelity is not sensitive to a global phase, hence it cannot distinguish operators $U$ and $-U$. Figure~\ref{fig:infid}(a) shows the average infidelities $\left(1-F\right)$ obtained for the different approaches as the level of noise increases. The GA and the NN always outperform the standard minimization routine. The advantage is also preserved in presence of moderate noise. Figure~\ref{fig:infid}(b) shows the standard deviation of the fidelity distributions for the three approaches with the same noise levels. Histograms in Figs.~\ref{fig:infid}(c)-(f) depict the infidelities distributions for different levels of noise. Surprisingly, we find that for a significant fraction ($\sim 20\%$) of synthetic processes the prediction resulting from the NMin routine is quite poor ($1-F>0.1$). We suspect that a similar behavior escaped previous numerical investigations, which is possibly emerging here because of the reduced number of input measurements that are processed, and the remarkably large set of randomly generated unitaries to be tested.


Besides the accuracy in retrieving arbitrary SU(2) processes, we analyze the time consumption of these algorithms \footnote{The algorithms are executed on a machine with an Intel Core i7-10700 CPU ($2.90$ GHz) and $32$ GB of RAM. The time consumption for the NN does not include the time required for training.}: on average, NMin requires $\sim 0.2 $~s for a single reconstruction, the GA requires $\sim 0.1 $ s, while, remarkably, the NN only requires $\sim 1 \,\mu$s.

As typical for non-deterministic algorithms, the predictions of the GA have been averaged over 10 runs for each unitary reconstruction. However, statistical fluctuations of the final prediction are marginal for all levels of noise. Accordingly, in the following applications of the GA, we will only consider single executions. 

\section{SPACE-DEPENDENT POLARIZATION TRANSFORMATIONS} \label{complex}
The techniques previously described address~{single} unitary polarization transformations that are spatially uniform, but can be extended to characterize space-dependent transformations $U(x,y)$. 
Spatially-inhomogeneous optical waveplates are such an example: they modify the transverse polarization profile of a collimated light beam, propagating along the $z$ axis of a reference frame. The action of a thin waveplate with its optic axis oriented at an angle $\alpha$ with respect to the $x$ axis can be put in the matrix form
\begin{equation}
R_{\delta}(x,y)=\begin{pmatrix}
\cos{(\delta/2)} & i\sin{(\delta/2)}e^{-2i \alpha(x,y)}\\
i\sin{(\delta/2)}e^{2i \alpha(x,y)} & \cos{(\delta/2)}
\end{pmatrix},
\label{eqn:complexwaveplate}
\end{equation}
where $\delta$ is the optical birefringence (that we assume to be uniform) and $\alpha (x,y)$ is space-dependent, that is, the optic axis takes different orientations across the transverse plane $(x,y)$. While the operator in Eq.\ \eqref{eqn:complexwaveplate} has no global phase factor, other optical devices could implement polarization transformations featuring a space-dependent global phase profile. To detect it, our technique should be accompanied by an interferometric measurement.
The complex transformation $U$ can be decomposed as a collection of local transformations, acting at different positions (pixels): 
\begin{equation}
    U=\sum_{(x,y)}U(x,y)\dyad{x,y}.
    \label{eqn:sumdecomposition}
\end{equation}
We generalize our algorithms to achieve the characterization of the full 2D maps by means of an iterative approach, i.e., via a pixel-by-pixel reconstruction.  

A straightforward iteration of the procedures described above could lead to some ambiguities, resulting in turn into the determination of discontinuous patterns $(\Theta(x,y),\bm{n}(x,y))$. This would be related to at least two issues. First, we cannot discriminate between processes $U$ and $-U$, leading to random jumps between the two families of solutions $(\Theta,\bm{n})$ and $(\pi-\Theta,-\bm{n})$. 
Second, physical imperfections in the sample or inaccuracies in the experimental procedures may lead to the reconstruction of an erroneous polarization transformation. To enforce a continuous modulation of the reconstructed parameters, we adopt different strategies, depending on the optimization algorithm. Motivated by these arguments, for each numerical technique (GA and NN) we identified a strategy to retrieve patterns that do not feature these artificial jumps.

\subsection{Continuity for the GA}

The imposition of the continuity constraint in the GA approach is performed \textit{a priori}, by properly preparing the starting population in each pixel. As long as the optical transformation does not feature singularities, we expect that the solution found in a given pixel cannot be too different from the solution associated with neighbouring pixels. Continuity can therefore be chased by initializing the population of a pixel from the solutions found by the GA in neighbouring pixels, possibly perturbed with uniform noise. 

\begin{figure}[t]
    \includegraphics[width=\linewidth]{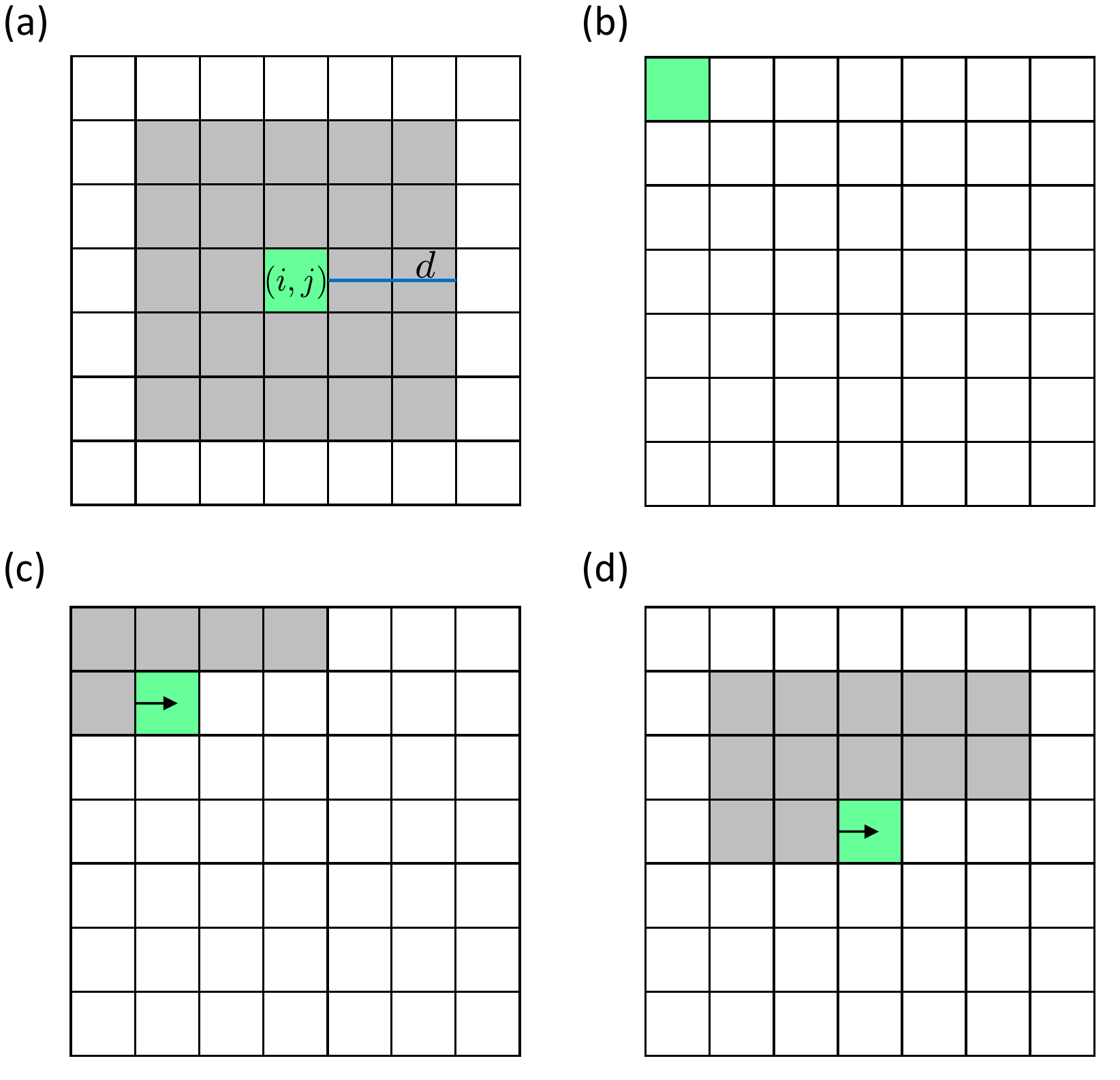}
    \caption{(a) Set of possible neighbouring pixels $\mathcal{S}$ (gray) for a certain grid position $(i,j)$ (green) within a distance ${d=2}$. (b)~The genetic optimization starts from pixel $(0,0)$, for which the initial population is completely random. Sets of neighbouring pixels used in our algorithm for pixels $(1,1)$ (c) and $(3,3)$ (d).}
    \label{fig:genetic_continuity}
\end{figure}

Formally, let us denote with $(i,j)$ the $i$-th row and $j$-th column of the pixel grid and with $s_{(i,j)}$ the best individual found by the GA performed in $(i,j)$. The initial population of the GA in $(i, j)$ is obtained by randomly selecting $N$ times (where $N$ is the population size) individuals from the set $\mathcal{S}$ of the best solutions found by the GA in all the neighbouring pixels within a distance $d$ from $(i, j)$ [see Fig.~\ref{fig:genetic_continuity}(a)].  Obviously, since the grid is spanned from left to right, the current pixel gets information only from neighbouring pixels where the GA has been already performed. As for the starting pixel $(0,0)$, no previous information is available, and the initial population is randomly generated [see Fig.\ \ref{fig:genetic_continuity}(b)]. Figures \ref{fig:genetic_continuity}(c)-(d) exemplify the choice of neighbours for pixels $(1,1)$ and $(3,3)$.
After performing $N$ samplings from the corresponding set $\mathcal{S}_{(i,j)}$ to prepare the initial population, a {uniformly-extracted} noise in the range $\left[ -\epsilon, \epsilon \right]$ is applied to each sampled solution. To respect the physical constraints, after the application of noise $\Theta$ is rescaled to the range $[ 0, \pi ]$, and the vector $\bm{n}$ is normalized. Finally, we observe that this procedure allows initializing the GA very close to the actual solution, so we expect fewer iterations to be needed to obtain an adequate convergence. Therefore, to decrease the computational time of the entire reconstruction, the number of generations for pixels other than $(0,0)$ is reduced. With a clever choice of the number of generations, this trick reduces the total time up to a factor of $10$. In particular, the parameter configuration chosen in our experiments is: $d=2$, $\epsilon =0.2$, $N_{\text{gen}}^{(0,0)} = 60$, $N_{\text{gen}}= 10$, where $N_{\text{gen}}^{(0,0)}$ indicates the number of iterations of the GA reconstructing the transformation associated with pixel $(0,0)$, while $N_{\text{gen}}$ refers to all the other pixels.  The remaining hyper-parameters are the same as those reported in Appendix \ref{app:hyper-parameters}.

\subsection{Continuity for the NN}
The imposition of the continuity constraint in the NN approach is performed \emph{a posteriori}, as feedforward networks do not possess a memory of previous inputs. After the transformation has been reconstructed pixel by pixel, the global phase ambiguity is gauged away by checking for sudden jumps in the predicted parameters $(\Theta(i,j),n_{x}(i,j),n_{y}(i,j))$ in correspondence with neighbouring pixels. If an inconsistency is found, the pixel parameters are switched to $\left(\pi - \Theta(i,j),-n_{x}(i,j),-n_{y}(i,j)\right)$ and, accordingly, $n_{z}(i,j)\rightarrow-n_{z}(i,j)$.



\section{Experimental results}
\begin{figure}[!t]
\includegraphics[width=0.5\textwidth]{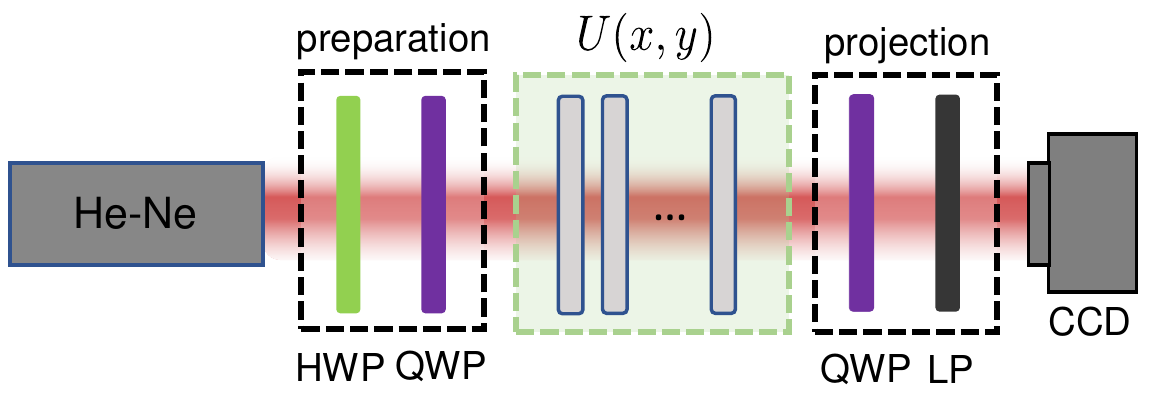}
\caption{Experimental setup to reconstruct complex polarization transformations, implemented via LC metasurfaces. The CCD camera records the light intensity distributions resulting from different projective measurements, realized by adjusting the preparation and projection stages.}
    \label{fig:setup}
\end{figure}

We realize complex polarization transformations with the setup sketched in Fig.~\ref{fig:setup}. We expand the light beam produced by a {He-Ne} laser source (with wavelength ${\lambda =633}$~nm), so as to have the final beam waist $w_0\simeq 5$~mm. To implement different realizations of Eq.~\eqref{eq:polar_mes}, we prepare the desired input polarization state with a half-wave plate (HWP) and a quarter-wave plate (QWP). The output projection is performed with a QWP and a linear polarizer (LP). Light intensity profiles are successively captured by a CCD camera placed immediately after the projection stage. 
In our implementation, the sum of Eq.~\eqref{eqn:sumdecomposition} extends over a discrete grid of $73\times 73$ pixels, obtained by compressing the experimental pictures captured by the camera. This allows minimizing the errors due to local intensity fluctuations in the image area, which appear when projecting onto different bases.\\ 
In our setup, the actual polarization transformation $U$ is realized via one or multiple LC metasurfaces, as those originally described in Refs.~\cite{Rubano2019,DErrico2020}. LC technology allows us to fabricate plates with a patterned local optic-axis orientation $\alpha(x,y)$, enabling us to implement several transformations within the same setup. In our experiments, we specifically employ linear polarization gratings, referred to as $g$-plates \cite{DErrico2020}, and LC plates with uniform optic-axis orientation, acting as ordinary waveplates. Furthermore, the birefringence $\delta$ of these devices is electrically controlled and can therefore be tuned \cite{Piccirillo2010}. In the case of a $g$-plate deflecting the beam along the $x$ direction, $\alpha(x,y)=\pi x/\Lambda$ ($\Lambda=5$ mm), while for ordinary waveplates we set $\alpha(x,y)=0$.

\begin{figure*}[!t]
     \centering
     \includegraphics[width=0.85\linewidth]{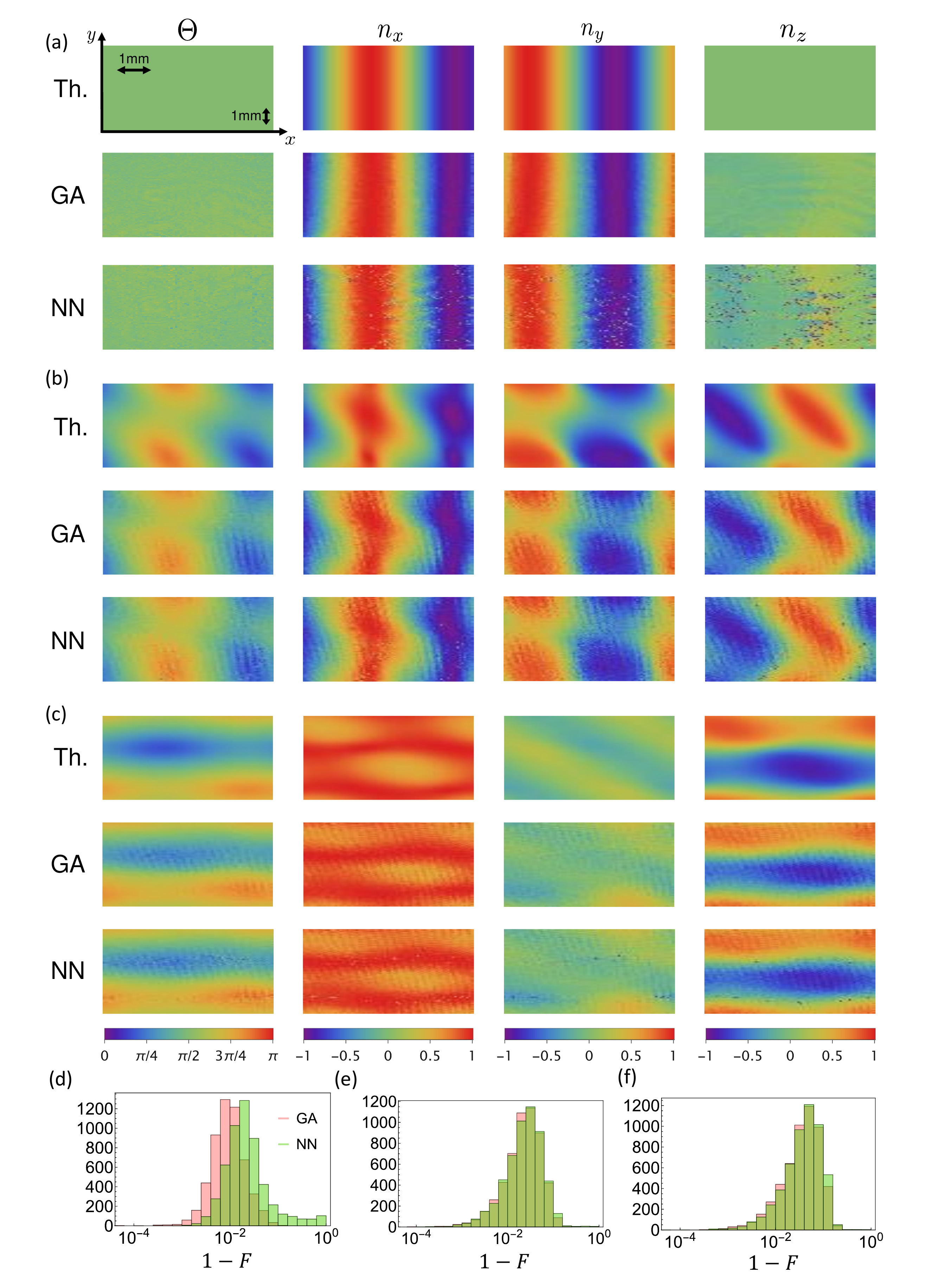}
     \caption{Reconstruction of the parameters $\left(\Theta, \bm{n} \right)$ associated with selected space-dependent polarization transformations: (a)~${U=T_x(\pi)}$, where $T_x$ denotes the $g$-plate operator acting along $x$, (b) ${U=T_y(\pi/4)T_x(\pi)W(\pi/2)}$, (c)~${U=T_y(\pi/2)T_x(\pi/6)W(\pi)}$. (d)-(f) Infidelities histograms for the pixels of processes (a)-(c), respectively. Computation time: (a) $106$ s (GA), $180$ ms (NN), (b) $109$ s (GA), $180$ ms (NN), (c) $106$ s (GA), $180$ ms (NN).} 
    \label{fig:processes}
\end{figure*}

To test the performances of our approaches when processing data from real experiments, we start implementing the simple polarization transformation realized by a single $g$-plate $T_x(\delta)$, oriented along the $x$ direction, tuned at $\delta=\pi$. Reconstructions obtained via GA and ML are compared with the theoretical parameters in Fig.\ \ref{fig:processes}(a). An excellent agreement is observed for this preliminary experiment: $\bar{F}_{GA}=98.6\%$ and $\bar{F}_{ML}=94.6\%$, 
where $\bar{F}$ denotes the average fidelity computed over all the pixels. In addition to the intrinsic precision limit of our routines, the non-perfect agreement has to be also ascribed to possible deviations of our setup from the ideal behaviour, especially to fabrication defects in our LC devices.  

We then proceed to characterize more complex polarization transformations, realized by stacking three metasurfaces. We specifically reconstruct the two processes ${U=T_y(\pi/4)T_x(\pi)W(\pi/2)}$ and ${U=T_y(\pi/2)T_x(\pi/6)W(\pi)}$ [see Figs.\ \ref{fig:processes}(b)-(c)], where $T_x$ ($T_y$) denotes the $g$-plate operator acting along the $x$ ($y$) direction, while $W(\delta)$ represents an ordinary waveplate. For $\delta=\pi/2$ and $\delta=\pi$, the latter acts as a QWP and HWP, respectively. In the first case, we find $\bar{F}_{GA}=97.0\%$ and $\bar{F}_{ML}=96.6\%$, 
while in the second realization we obtain $\bar{F}_{GA}=95.3\%$ and $\bar{F}_{ML}=94.7\%$. 
The times required for the full reconstruction are indicated in the captions for each process. While we typically get a lower fidelity in the case of the NN, this method is about $600$ times faster than the one based on the GA. Histograms in Figs.~\ref{fig:processes}(d)-(f) depict the infidelities distributions over all the pixels for the three processes described above.  \\
The high values of the recorded average fidelities confirm that our routines are suitable for reconstructing optical unitary operations from a very limited amount of experimental data. The good performances of these tomographic approaches in the case of a real experimental scenario also represent a crucial demonstration of their robustness against realistic sources of noise in the detection apparatus, which are more general and less controllable than our synthetic estimates. Finally, we observe that only in Fig.\ \ref{fig:processes}(a) the NN fails to reconstruct the correct values of $(n_x,n_y,n_z)$ at isolated pixels, while this is not taking place in Figs.\ \ref{fig:processes}(b)-(c). As such, we believe that the imperfections in Fig.\ \ref{fig:processes}(a) are strictly related to the special case $n_z=0$.

\section{CONCLUSIONS AND OUTLOOK}
We presented two routines which can assist experimentalists in characterizing optical setups used to manipulate light polarization. When adopting a GA and a NN, we obtain noteworthy performances in reconstructing complex polarization transformations from a set of input measurements close to the minimal case, both in terms of the accuracy of the reconstruction and the required computational resources. In future implementations, these approaches could be used in a complementary way. ML appears more fitting for a preliminary scanning of the experimental platform, while the GA produces the most reliable predictions, even if more time is required to converge to the correct solution.  
We are currently planning to optimize these routines even further. For instance, ML predictions could substantially improve by processing the whole experimental images at once, by means of convolutional neural networks \cite{o2015introduction,notesMarquardt}. This approach would facilitate the reconstruction of the optical process even in presence of singularities, and without relying upon post-processing adjustments, as we did in this work.
The final predictions might be further processed to retrieve geometrical and topological features of unitary evolutions which can be simulated within our photonic setup \cite{cardano2017detection,DErrico2020,DiColandrea2023,Yi2023}.
At the same time, it would be interesting to extend the same approaches to more general physical scenarios, for example exploring non-unitary evolutions~\cite{PhysRevLett.119.130501} and de-polarizing channels, probing multi-photon regimes~\cite{zhong2020quantum} in quantum photonics experiments, or employing these techniques for biomedical applications based on polarization optics \cite{He2021}. 
\section*{ACKNOWLEDGMENTS}
FDC and FC acknowledge financial support from the European Union Horizon 2020 program, under European Research Council (ERC) Grant No. 694683 (PHOSPhOR). FC acknowledges financial support from PNRR MUR project PE0000023-NQSTI. LA acknowledges financial support from the Swiss National Science Foundation (SNSF) under grant No. 204801021001. A.D. acknowledges financial support from ERC AdG NOQIA; Ministerio de Ciencia y Innovation Agencia Estatal de Investigaciones (PGC2018-097027-B-I00/10.13039/501100011033,  CEX2019-000910-S/10.13039/501100011033, Plan National FIDEUA PID2019-106901GB-I00, FPI, QUANTERA MAQS PCI2019-111828-2, QUANTERA DYNAMITE PCI2022-132919,  Proyectos de I+D+I “Retos Colaboración” QUSPIN RTC2019-007196-7); European Union NextGenerationEU (PRTR);  Fundació Cellex; Fundació Mir-Puig; Generalitat de Catalunya (European Social Fund FEDER and CERCA program (AGAUR Grant No. 2017 SGR 134, QuantumCAT \ U16-011424, co-funded by ERDF Operational Program of Catalonia 2014-2020); Barcelona Supercomputing Center MareNostrum (FI-2022-1-0042); EU Horizon 2020 FET-OPEN OPTOlogic (Grant No 899794); National Science Centre, Poland (Symfonia Grant No. 2016/20/W/ST4/00314); European Union’s Horizon 2020 research and innovation programme under the Marie-Skłodowska-Curie grant agreement No 101029393 (STREDCH) and No 847648  (“La Caixa” Junior Leaders fellowships ID100010434: LCF/BQ/PI19/11690013, LCF/BQ/PI20/11760031,  LCF/BQ/PR20/11770012, LCF/BQ/PR21/11840013).  A.D. further acknowledges the financial support from a fellowship granted by la Caixa Foundation (ID 100010434, fellowship code LCF/BQ/PR20/11770012).

\section*{Data and Code availability statement}
The source code and data used to produce the results and analyses presented in this manuscript are available from GitHub repository: \href{https://github.com/1234534253/QPT}{https://github.com/1234534253/QPT}

\section*{Conflicts of interest}
The authors declare no conflict of interest.

------------
\clearpage
\appendix

\section{Minimal set of measurements}
\label{app:supp_proof}
In this section, we provide two arguments explaining why five is the minimal number of projective measurements needed to reconstruct a generic SU(2) operator.

We first start with a geometric argument. Reconstructing a generic $SU(2)$ operator $U$ means predicting its action on any qubit. Accordingly, given two different input states, we need to determine the obtained output states. Let us assume that our first input state is $\ket{\psi_{\text{in},1}}$. The output state $U\ket{\psi_{\text{in},1}}$ is a point on the Bloch sphere. To unambiguously identify a point on a sphere, the distance from three other points must be determined (as long as these do not lie on a great circle). ``Measuring distances" is equivalent to extracting the magnitude of the scalar product. Therefore, we perform three projective measurements of the state $U\ket{\psi_{\text{in},1}}$. In this way, we fully reconstruct $U\ket{\psi_{\text{in},1}}$. We now choose a second input state $\ket{\psi_{\text{in},2}}$ that, crucially, is not orthogonal to the first one. Accordingly, to reconstruct $U\ket{\psi_{\text{in},2}}$ only two projective measurements are needed, as the knowledge of the distance from $U\ket{\psi_{\text{in},1}}$ is already given by the unitarity condition.

This gives us five measurements. However, when reconstructing the output state, it may happen that $U\ket{\psi_{\text{in},1}}$ and the two other states chosen as reference points all lie on a great circle on the Bloch sphere. There is actually only a zero-measure set of states for which this may happen. However, with data coming from real experiments, this possibility compromises the robustness of the reconstruction. This effect can be tackled by introducing a sixth measurement, which always ensures an accurate reconstruction.

We also provide an algebraic argument. A generic SU(2) operator $U$ can be put in the matrix form
\begin{equation}
U=
\begin{pmatrix}
A e^{i \varphi} && Be^{i \psi}\\
-B e^{-i \psi} && A e^{-i\varphi}
\end{pmatrix},
\end{equation}
where $A$ and $B$ are positive real numbers satisfying ${A^2+B^2=1}$, and $\varphi$ and $\psi$ represent real phases. 

In the basis of circular polarizations, we can determine both $A$ and $B$ with a single projective measurement:
\begin{equation}
I_{LL}=\abs{\mel{L}{U}{L}}^2=A^2.
\end{equation}
From the above equation, we determine $A=\sqrt{I_{LL}}$ and $B=\sqrt{1-I_{LL}}$. 

As for the phases $\varphi$ and $\psi$, four additional measurements are required.
We consider the combination
\begin{equation}
\begin{split}
I_{LH}&=\abs{\mel{H}{U}{L}}^2=\dfrac{1}{2}-AB \cos\left(\varphi+\psi \right),\\
I_{LD}&=\abs{\mel{D}{U}{L}}^2=\dfrac{1}{2}+AB \sin\left(\varphi+\psi \right),
\end{split}
\end{equation}
from which we can \emph{uniquely} determine the phase sum $\varphi+\psi$, and
\begin{equation}
\begin{split}
I_{HL}&=\abs{\mel{L}{U}{H}}^2=\dfrac{1}{2}+AB \cos\left(\varphi-\psi \right),\\
I_{HD}&=\abs{\mel{D}{U}{H}}^2=\dfrac{1}{2}+A^2 \sin\left(2\varphi\right) + B^2\sin\left(2\psi\right) ,
\end{split}
\end{equation}
from which we can \emph{uniquely} determine the phase difference $\varphi-\psi$.

\section{Genetic tomography}
\label{app:GAtomography}
Here follows the detailed sequence of operators used in our GA. First, the well-known \emph{tournament selection} mechanism \cite{goldberg1991comparative} is used as selection operator. This consists in repeating $N$ times (where $N$ is the population size) the following steps: 
\begin{enumerate}
    \item randomly select a subset of $k$ individuals;
    \item choose the fittest individual among them to be inserted in the 
    \emph{mating pool}, where reproduction takes place.
\end{enumerate} 
For our purposes, ``fittest" are those individuals for which the cost function is minimum. 
Second, the blend crossover \cite{eshelman1993real} is used to mate individuals in the mating pool. Accordingly, when two individuals ${\bm{x}_A=(\Theta_A,n_{xA},n_{yA},n_{zA})}$ and ${\bm{x}_B=(\Theta_B,n_{xB},n_{yB},n_{zB})}$ reproduce, two newborn individuals $\bm{x}_1$ and $\bm{x}_2$ originates, with each \emph{gene} $x_{c,i}$ given by a random number belonging to the interval
\begin{equation}
[x_{A,i}-\alpha_c (x_{B,i}-x_{A,i}),x_{B,i}+\alpha_c (x_{B,i}-x_{A,i})],
\label{eqn:reproduction}
\end{equation}
where $\alpha_c$ tunes the crossover, and we are assuming ${x_{B,i}\geq x_{A,i}}$, with $c=\left\lbrace 1,2\right\rbrace$ and $i=\left\lbrace 1,2,3,4 \right\rbrace$.
Finally, a Gaussian mutation \cite{kramer2017genetic} with mean $\mu$ and standard deviation $\sigma$ can mutate individual genes of offspring solutions.
Our GA is also provided with an \emph{elitism} mechanism. In other words, the best chromosome from the old population is carried over to the next one, replacing the worst individual of the offspring. This mechanism pushes the algorithm to a fast convergence toward the best solution. 
To preserve the physical validity of the final prediction, after each operation carried out on an individual, a modulo-$\pi$ operation is performed on its first gene, while normalization is enforced on its last three genes.
The maximum number of generations is used as a termination criterion.

Table \ref{ga_code} reports the pseudo-code of the implemented GA.  Our GA is performed relying on the \texttt{DEAP} library~\cite{DEAP_JMLR2012}.

\begin{table}[h!]
\caption{Pseudo-code of the implemented genetic algorithm.}
\centering
\vspace{10pt}
\scriptsize
\begin{algorithmic}[1]
\hrule
\vspace{1.5pt}
\hrule
\medskip
\REQUIRE size of the population $pop\_size$, $k$ for tournament selection, crossover probability $p_c$, $\alpha$ for blend crossover, mutation probability $p_m$, $\mu$ and $\sigma$ for Gaussian mutation, termination criterion $t$.
\medskip
\ENSURE the best solution $best$.
\medskip
\STATE $gen \leftarrow 0$;
\STATE $pop \leftarrow$ generateRandomPopulation($pop\_size$); \\
\STATE checkPhysicalConstraints($\textit{pop}$);
\\
\STATE evaluateFitness($pop$); 
\STATE $best \leftarrow$ getBestIndividual($pop$); 
\WHILE{($t$ is not satisfied)} 
\STATE $\textit{offspring} \leftarrow$ executeTournament($pop$, $k$); \\
\STATE executeBlendCrossover($\textit{offspring}$, $p_c$, $\alpha$); \\
\STATE checkPhysicalConstraints($\textit{offspring}$)
\\
\STATE executeGaussianMutation($\textit{offspring}$, $p_m$, $\mu$, $\sigma$); \\
\STATE checkPhysicalConstraints($\textit{offspring}$)\\
\STATE evaluateFitness($\textit{offspring}$); \\
\STATE $pop \leftarrow \textit{offspring}$; \\
\STATE $pop \leftarrow$ elitism($pop$, $best$); \\
\STATE $best \leftarrow$ getBestIndividual($pop$);\\ 
\STATE $gen \leftarrow gen+1$;\\
\ENDWHILE
\RETURN $best$;
\medskip
\hrule
\vspace{1.5pt}
\hrule
\end{algorithmic}
\label{ga_code}
\end{table}

\section{Machine Learning tomography}
\label{app:MLtomography}
\begin{figure}[t]
\centering
\includegraphics[width=\linewidth]{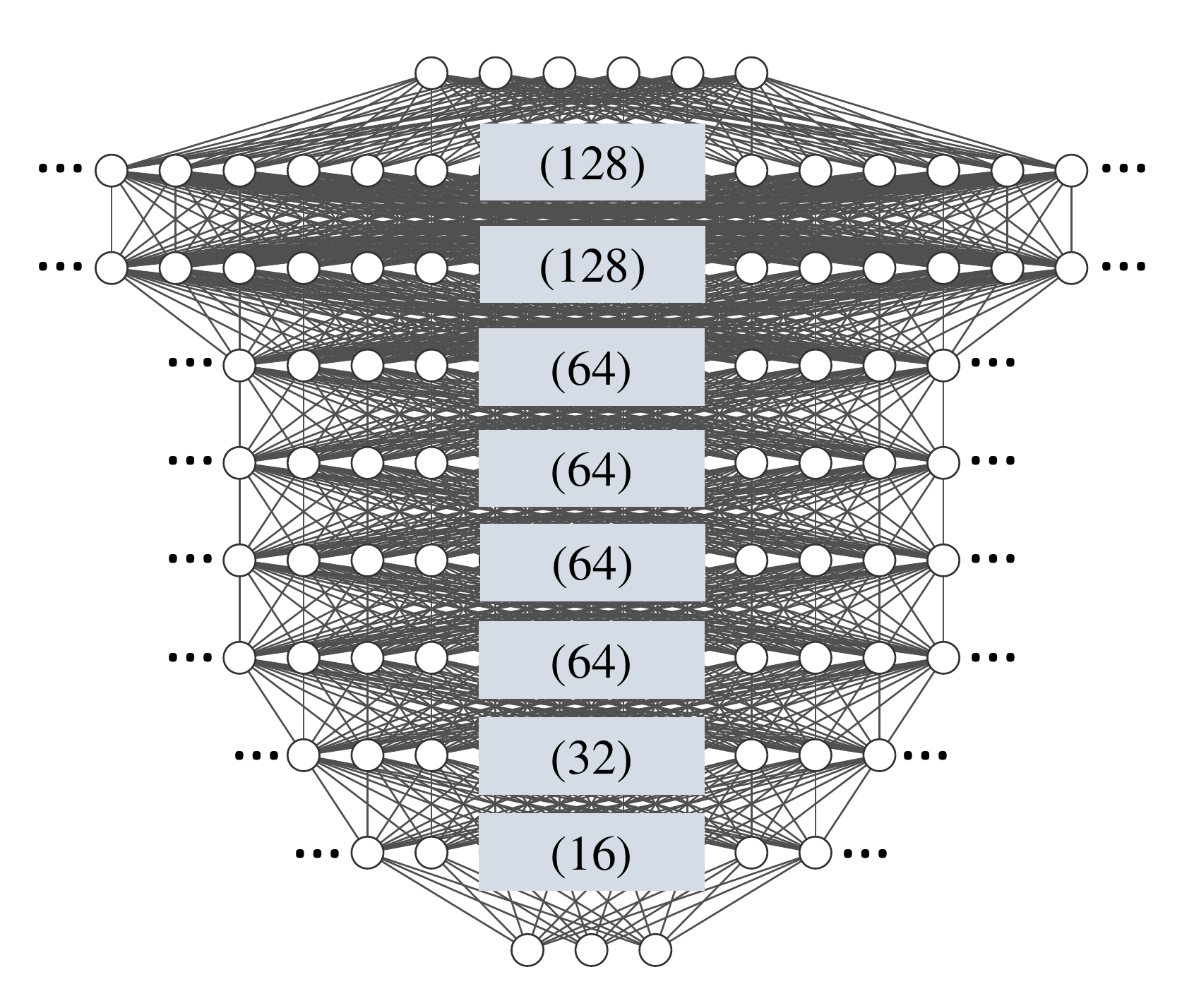}
\caption{\label{fig:figML}
Structure of the dense neural network. For the intermediate layers (\emph{hidden layers}),
the number of neurons has been written explicitly.
}
\end{figure}

The network scheme we consider is a \emph{dense} network, as each neuron in a given layer is connected to each neuron in the subsequent layer. Figure~\ref{fig:figML} shows the structure of our network. We choose the rectified linear unit (``ReLu") as activation function for all the layers, except for the output layer, for which we choose the Sigmoid function.
Training data are generated at run-time, so as to avoid importing huge data-sets and possibly decreasing the issue of \emph{overfitting}, which can result from a too limited number of training samples. Several standard loss functions have been tested for the supervised learning. Among these, the MSE has always shown the fastest convergence. 
The learning process is divided into $50$ epochs. During each epoch, the NN learns about a training set of $2^{20}$ randomly generated SU(2) processes, divided into $2^{12}$ \emph{batches}. At the end of each epoch, the performance of the network is evaluated on a different set of data, the \emph{validation} set, consisting of $2^{18}$ processes, divided into $2^{10}$ batches. The validation phase works similarly to the training, but after the evaluation of the cost function the weights are not adjusted. To improve the learning performance, we employ the callback ``ReduceLROnPlateau". This function allows us to keep track of the loss during the various epochs and reduce the learning rate in case of stagnation. To ensure good performances in case of noisy data, we also employ Gaussian Dropout, where noise is randomly applied to chosen layers during training~\cite{labach2019survey,JMLR:v15:srivastava14a}.  Finally, the Adam optimization algorithm \cite{kingma2017adam} is used to adjust the network hyper-parameters during the supervised learning. 

The ML approach is implemented with the help of the \texttt{TENSORFLOW} library \cite{tensorflow2015-whitepaper``}. 

\section{Hyper-parameters}
\label{app:hyper-parameters}
For the sake of reproducibility, we declare the configurations employed in our approaches. 
After a careful tuning procedure, the best setting of the GA and NN hyper-parameters is reported in Table \ref{tab:GA_parameters} and \ref{tab:NN_parameters}, respectively.

\begin{table}[!h]
\caption{GA hyper-parameters}
\centering
\begin{tabular}{ll}
\multicolumn{2}{c}{}
\\\hline
\hline
\\
Population size       & $N = 40$                     \\
Number of generations & $N_\text{gen}$ = 60               \\ 
\\\hline
\multicolumn{2}{c}{Selection parameters}     
\\ \hline
\\
Tournament size       & $k = 3$     
\\
\\ \hline
\multicolumn{2}{c}{Crossover parameters}           \\ \hline
\\
Crossover probability & $p_c$ = 0.8                \\
Blend crossover       & $\alpha_1=\alpha_2= 0.5$      
\\
\\ \hline
\multicolumn{2}{c}{Mutation parameters}            \\ \hline
\\
Mutation probability  & $p_m$ = 0.1                \\
Gaussian mutation     & $\mu$ = 0, $\sigma$ = 0.2 \\
\\
\hline
\hline
\end{tabular}
\label{tab:GA_parameters}
\end{table}

\begin{table}[!h]
\caption{NN hyper-parameters}
\label{tab:NN_parameters}
\centering
\begin{tabular}{ll}
\multicolumn{2}{c}{}
\\\hline \hline
\\
Number of hidden layers       & $N =  8$                   \\
Neurons in input layer & $N_\text{in}=6$ \\
Neurons in layer $1$ & $N_1$ = 128               \\ 
Neurons in layer $2$ & $N_2$ = 128               \\ 
Neurons in layer $3$ & $N_3$ = 64             \\ 
Neurons in layer $4$ & $N_4$ = 64             \\ 
Neurons in layer $5$ & $N_5$ = 64             \\ 
Neurons in layer $6$ & $N_6$ = 64             \\
Neurons in layer $7$ & $N_7$ = 32             \\ 
Neurons in layer $8$ & $N_8$ = 16             \\ 
Neurons in output layer & $N_\text{out}$ = 3             \\ 
\\\hline
\multicolumn{2}{c}{Activations}     
\\ \hline
\\
Layers 1, 2, 3, 4, 5, 6, 7, 8      & \qo{ReLu}  \\
Output layer & \qo{Sigmoid}\\

\\ \hline
\multicolumn{2}{c}{Data structure}           \\ \hline
\\
Batch size   & $S_{batch}=256$\\
Batches per epoch (training) & $N_{T}=2^{12}$\\
Batches per epoch (validation) & $N_{V}=2^{10}$\\
Number of epochs & $N_E=50$
\\
\\ \hline
\multicolumn{2}{c}{Optimization parameters}           \\ \hline
\\
Optimizer   & \qo{Adam}\\
Loss function & MSE\\
Learning rate (initial) & $\eta=10^{-3}$\\
Gaussian Dropout rate & $\text{rate}=0.01$\\
ReduceLROnPlateau:\\
Patience & $5$ epochs\\
Reduction factor & $\mu=0.1$
\\  \\ \hline \hline

\end{tabular}

\end{table}

\clearpage
\bibliography{quantum_process_tomography}

\end{document}